\documentstyle[prl,aps]{revtex}

\begin{document}
\author{Jian-Qi Shen$^{1}$\footnote{E-mail address: jqshen@coer.zju.edu.cn}, Zhi-Chao Ruan$^{1}$,
and Sailing He$^{1,2}$\footnote{E-mail address: sailing@kth.se}}
\address{$^1$ Centre for Optical
and Electromagnetic Research, Joint Research Centre of Photonics
of the Royal Institute of Technology (Sweden) and Zhejiang
University, Zhejiang University, Hangzhou Yuquan 310027, P. R.
China\\
$^2$ Laboratory of Photonics and Microwave Engineering,
Department of Microelectronics and Information Technology,\\
Royal Institute of Technology, Electrum 229, SE-164 40 Kista,
Sweden}
\date{\today }
\title{How to realize a negative refractive index material at the atomic level \\in an optical frequency range}
\maketitle

\begin{abstract}
The theoretical mechanism for realizing the negative refractive
index with {\it electromagnetically induced transparency} (EIT) is
studied. It is shown that in a three-level dense atomic gas, there
is a frequency band in which the EIT medium will exhibit
simultaneously negative electric permittivity and magnetic
permeability in the optical frequency range, and the atomic gas
thus becomes a left-handed material. The expressions for the
electric permittivity and the magnetic permeability for the probe
frequency is presented. One of the remarkable features of the
present novel scheme is such that the obtained EIT left-handed
material is {\it isotropic} and may therefore have some
potentially important applications in the development of new
techniques in quantum optics.
\end{abstract}
\section{INTRODUCTION}
More recently, a kind of artificial composite metamaterials
(so-called {\it left-handed media} or {\it negative refractive
index media}) having a frequency band where the electric
permittivity and the magnetic permeability are simultaneously
negative attracts considerable attention of many researchers in
various fields such as materials science, condensed matter
physics, optics and classical applied electromagnetism
\cite{Veselago,Smith,Shelby,Pendry2,Pendry1,Shelby2}. It can be
readily verified that the left-handed media exhibit a number of
peculiar electromagnetic and optical properties, including
reversals of both the Doppler shift and Cherenkov radiation,
anomalous refraction and amplification of evanescent wave. In
experiments, a combination of two structures ({\it array of long
metallic wires} and {\it split ring resonators}
\cite{Pendry2,Pendry1}) yields such a type of artificial negative
refractive index media \cite{Shelby}. Although Veselago's original
paper \cite{Veselago} and most of the recent theoretical works
investigated mainly the electromagnetic and optical properties in
the {\it isotropic} left-handed media\cite{Shen}, up to now, the
left-handed media that have been prepared successfully
experimentally are actually {\it anisotropic} in nature, and it
may be very difficult to prepare an isotropic left-handed medium
\cite{Smith,Shelby2,Hu}. In the previous experimental and
theoretical work, investigators used classical optical approaches
to design and fabricate the negative refractive index
materials\cite{Shelby,Pendry2,Pendry1,Chen,Simo}. In the present
paper, we will suggest an alternative method (quantum optical
approach) to realize the negative refractive index: specifically,
under certain conditions, the electric-dipole and magnetic-dipole
transitions in a multilevel EIT (electromagnetically induced
transparency) atomic system will exhibit simultaneously negative
permittivity and negative permeability. Recently, many theoretical
and experimental investigations show that the control of phase
coherence in multilevel atomic ensembles will give rise to many
novel and striking quantum optical phenomena in the wave
propagation of near-resonant light\cite{Harris,Zhusy,Zhuy2}. One
of the most interesting phenomena is electromagnetically induced
transparency\cite{Harris}. More recently, some unusual physical
effects associated with EIT observed experimentally include the
ultraslow light pulse propagation, superluminal light propagation,
light storage in atomic vapor and atomic ground state cooling,
some of which are believed to be useful for the development of new
techniques in quantum optics\cite{Harris,Li,Schmidt}.

In the following, we will consider a new property of atomic media,
{\it i.e.}, the possibility for the EIT media to become the
left-handed media under some conditions. Consider a $\Lambda$-type
three-level atomic ensemble with one upper level $|a\rangle$ and
two lower levels $|b\rangle$ and $|c\rangle$ (see Fig. 1). Such an
atomic system interacts with two optical fields, {\it i.e.}, the
coupling laser and the probe laser, which couple the level pairs
$|a\rangle$-$|c\rangle$ and $|a\rangle$-$|b\rangle$, respectively.
Here we assume that the coupling laser is in resonance with the
$|a\rangle$-$|c\rangle$ transition, while the probe laser has a
frequency detuning $\Delta$ that is defined by
$\Delta=\omega_{ab}-\omega$, where $\omega_{ab}$ and $\omega$
denote the $|a\rangle$-$|b\rangle$ transition frequency and the
probe mode frequency, respectively. Note that since the level
pairs $|a\rangle$-$|c\rangle$ and $|a\rangle$-$|b\rangle$ can be
coupled to two laser fields, the parity of level $|a\rangle$ is
different from both $|b\rangle$ and $|c\rangle$. If, for example,
$|a\rangle$ possess an odd parity, $|b\rangle$ and $|c\rangle$
will have even parity. Thus, the electric dipole matrix elements
$\vec{\wp}_{ab}=\langle a|e\vec{r}|b\rangle\neq 0$ and
$\vec{\wp}_{cb}=\langle c|e\vec{r}|b\rangle=0$, and the magnetic
dipole matrix elements $\vec{m}_{cb}=\langle c|(e/2m_{\rm
e})(\vec{L}+2\vec{S})|b\rangle \neq 0$ and $\vec{m}_{ab}=\langle
a|(e/2m_{\rm e})(\vec{L}+2\vec{S})|b\rangle=0$, where $\vec{L}$
and $\vec{S}$ denote the operators of the orbital angular momentum
and spin of electrons, respectively. So, it is possible for the
nearly resonant probe laser to cause the electric-dipole
transition between levels $|a\rangle$ and $|b\rangle$, and the
magnetic-dipole transition between levels $|c\rangle$ and
$|b\rangle$ in the three-level atomic medium. The electric-dipole
transition ($|a\rangle$-$|b\rangle$) and the magnetic-dipole
transition ($|c\rangle$-$|b\rangle$) will yield the electric
polarizability and the magnetic susceptibility at probe frequency,
respectively. In general, the dimensionless ratio
$|\vec{m}_{cb}/\left(\vec{\wp}_{ab}c\right)|\simeq 10^{-2}$ in an
atomic system, where $c$ denotes the speed of light in a vacuum.
For this reason, the magnetic-dipole transition may not be
considered in the treatment for the wave propagation in an
artificially electromagnetic material. However, in a three-level
EIT medium where the intensity of coupling laser is much larger
than that of the probe light, the population in level $|c\rangle$
is much greater than that in the upper level $|a\rangle$. In other
words, the stronger coupling laser enhances the probability
amplitude of level $|c\rangle$. Thus, the order of magnitude of
the density matrix element $\rho_{cb}$ may be larger than that of
$\rho_{ab}$. This, therefore, means that the magnetic dipole
moment ($2{m}_{cb}^{\ast}\rho_{cb}$) may possibly have the same
order of magnitude of the electric dipole moment
($2c{\wp}_{ab}^{\ast}\rho_{ab}$). Further analysis shows that such
an EIT medium may exhibit negative permittivity and permeability,
and will therefore become an ideal candidate for realizing
isotropic left-handed media.
\section{PERMITTIVITY AND PERMEABILITY IN A THREE-LEVEL EIT medium}
We should first consider the steady density matrix elements of
this EIT. The density matrix elements, $\rho_{ab}$ and
$\rho_{cb}$, of such a three-level system can be rewritten as
$\rho_{ab}=\tilde{\rho}_{ab}\exp\left(-i\omega t\right)$ and
$\rho_{cb}=\tilde{\rho}_{cb}\exp\left[-i\left(\omega+\omega_{ca}\right)t\right]$.
Here, we assume that the intensity of the probe laser is
sufficiently weak and therefore nearly all the atoms remain in the
ground state, {\it i.e.}, the atomic population in level
$|b\rangle$ is unity. Under this assumption, $\tilde{\rho}_{ab}$
and $\tilde{\rho}_{cb}$ satisfy the following matrix equation
\cite{Scully}
\begin{equation}
\frac{\partial}{\partial t}\left( {\begin{array}{*{20}c}
   {\tilde{\rho}_{ab}}  \\
   {\tilde{\rho}_{cb}}  \\
\end{array}} \right)=\left( {\begin{array}{*{20}c}
   {-\left(\gamma_{1}+i\Delta\right)} & {\frac{i}{2}\Omega_{\rm c}}  \\
   {\frac{i}{2}\Omega_{\rm c}^{\ast}} & {-\left(\gamma_{3}+i\Delta\right)}  \\
\end{array}} \right)\left( {\begin{array}{*{20}c}
    {\tilde{\rho}_{ab}}  \\
   {\tilde{\rho}_{cb}}  \\
\end{array}} \right)+\left( {\begin{array}{*{20}c}
    {\frac{i\wp_{ab}{\mathcal E}}{2\hbar}}  \\
   {0}  \\
\end{array}} \right)
\label{density},
\end{equation}
where $\gamma_{1}$ and $\gamma_{3}$ represent the spontaneous
decay rate of level $|a\rangle$ and the dephasing rate
(nonradiative decay rate) of $|c\rangle$, respectively.
$\wp_{ab}$, ${\mathcal E}$ and $\Omega_{\rm c}$ denote the
electric dipole matrix element, the probe field envelope and the
Rabi frequency of coupling laser ($\Omega_{\rm c}=\wp_{ac}{E}_{\rm
c}/{\hbar}$ with ${E}_{\rm c}$ being the electric field strength
of the coupling laser). It can be readily verified that the steady
solution of Eq. (\ref{density}) takes the following form
\begin{equation}
\tilde{\rho}_{ab}=\frac{i\wp_{ab}{\mathcal
E}\left(\gamma_{3}+i\Delta\right)}{2\hbar\left[\left(\gamma_{1}+i\Delta\right)\left(\gamma_{3}+i\Delta\right)+\frac{\Omega_{\rm
c}^{\ast}\Omega_{\rm c}}{4}\right]},      \qquad
\tilde{\rho}_{cb}=-\frac{\wp_{ab}{\mathcal E}\Omega_{\rm
c}^{\ast}}{4\hbar
\left[\left(\gamma_{1}+i\Delta\right)\left(\gamma_{3}+i\Delta\right)+\frac{\Omega_{\rm
c}^{\ast}\Omega_{\rm c}}{4}\right]}.      \label{steadysolution}
\end{equation}
Apparently, there exists a relation between $\tilde{\rho}_{cb}$
and $\tilde{\rho}_{ab}$, {\it i.e.},
\begin{equation}
\tilde{\rho}_{cb}=\frac{i}{2}\left(\frac{\Omega_{\rm
c}^{\ast}}{\gamma_{3}+i\Delta}\right)\tilde{\rho}_{ab}.
\end{equation}

As to the problem of local field correction, here, in the simplest
case (valid for gases), we can take the local field to be the same
as the macroscopic field (average field in the sample). So, we
need not consider the Clausius-Mossotti-Lorentz relation
\cite{Jackson}. In a three-level atomic system, the electric
polarizability $\chi_{\rm e}$ and the magnetic susceptibility
$\chi_{\rm m}$ are of the form $\chi_{\rm
e}=2N\wp_{ab}^{\ast}\tilde{\rho}_{ab}/\left(\epsilon_{0}{\mathcal
E}\right)$ and $\chi_{\rm
m}=2Nm_{cb}^{\ast}\tilde{\rho}_{cb}/{\mathcal H}$, respectively,
where $N$ denotes the atomic density (total number of atoms per
volume). Thus, with the help of the steady solution
(\ref{steadysolution}), one can obtain the relative permittivity
$\epsilon_{\rm r}=1+\chi_{\rm e}$ and the relative permeability
$\mu_{\rm r}=1+\chi_{\rm m}$ of the above atomic system. By using
the relation ${\mathcal H}=\sqrt{\epsilon_{\rm
r}\epsilon_{0}/\mu_{\rm r}\mu_{0}}{\mathcal E}$ between the
envelopes of magnetic and electric fields, we have the expressions
for the electric polarizability and the magnetic susceptibility,
{\it i.e.},
\begin{equation}
\left\{
\begin{array}{ll}
& \chi_{\rm
e}=\frac{iN|\wp_{ab}|^{2}\left(\gamma_{3}+i\Delta\right)}{\epsilon_{0}\hbar\left[\left(\gamma_{1}+i\Delta\right)\left(\gamma_{3}
+i\Delta\right)+\frac{\Omega_{\rm
c}^{\ast}\Omega_{\rm c}}{4}\right]},     \\
&   \chi_{\rm
m}=\frac{m_{cb}^{\ast}}{\wp_{ab}^{\ast}c}\sqrt{\frac{\mu_{\rm
r}}{\epsilon_{\rm r}}}\left(\frac{i}{2}\frac{\Omega_{\rm
c}^{\ast}}{\gamma_{3}+i\Delta}\right)\chi_{\rm e}.
\end{array}
\right. \label{expressions}
\end{equation}
Since the dephasing rate $\gamma_{3}$ is small compared with the
spontaneous decay rate $\gamma_{1}$ ($\gamma_{3}$ is in general
two or three orders of magnitude less than $\gamma_{1}$)\cite{Li},
in the following analysis we will ignore $\gamma_{3}$.

With the help of Eqs. (\ref{expressions}), one can obtain
\begin{equation}
\left\{
\begin{array}{ll}
&   \epsilon_{\rm r}=1+\chi_{\rm
e}=1-\frac{N|\wp_{ab}|^{2}\Delta}{\epsilon_{0}\hbar\left[i\Delta\left(\gamma_{1}+i\Delta\right)+\frac{\Omega_{\rm
c}^{\ast}\Omega_{\rm c}}{4}\right]},   \\
&  \mu_{\rm r}^{\pm}=1+\frac{\varsigma^{2}\pm
\sqrt{\varsigma^{4}+4\varsigma^{2}}}{2}  \quad
\left(\varsigma=\frac{m_{cb}^{\ast}}{\wp_{ab}^{\ast}c}\frac{\Omega_{\rm
c}^{\ast}}{2\Delta}\frac{\chi_{\rm e}}{\sqrt{1+\chi_{\rm e}}}
\right).
\end{array}
\right. \label{expressions2}
\end{equation}
In an EIT left-handed medium, if $\chi_{\rm e}$ has a very small
imaginary part, the parameter $\varsigma$ will be an imaginary
number. It follows from Eqs. (\ref{expressions2}) that if
$\varsigma^{2}<-4$, the magnetic permeability $\mu_{\rm r}$ will
have a negative real part and a nearly vanishing imaginary part.
Further analysis shows that both the negative and the positive
roots ($\chi_{\rm m}^{\pm}=(\varsigma\pm
\sqrt{\varsigma^{4}+4\varsigma^{2}})/2$) of the magnetic
susceptibility are valid for Eqs. (\ref{expressions}).

In the following section, we will find out a frequency band in
which the three-level EIT atomic gas will exhibit simultaneously
negative electric permittivity and magnetic permeability.

\section{EXISTENCE OF NEGATIVE PERMITTIVITY AND PERMEABILITY}

In order to find out the conditions for realizing the negative
optical indices, we should first analyze the real and imaginary
parts of the electric polarizability $\chi_{\rm e}$ in Eqs.
(\ref{expressions2}). Set $\chi_{\rm e}=\chi'_{\rm e}+i\chi''_{\rm
e}$. Then by the aid of Eq. (\ref{expressions}), one can arrive at
\begin{equation}
\chi'_{\rm e}=\frac{N|\wp_{ab}|^{2}\Delta
\left(\Delta^{2}-\frac{\Omega_{\rm c}^{\ast}\Omega_{\rm
c}}{4}\right)}{\epsilon_{0}\hbar\left[\left(\Delta^{2}-\frac{\Omega_{\rm
c}^{\ast}\Omega_{\rm
c}}{4}\right)^{2}+\Delta^{2}\gamma_{1}^{2}\right]},    \qquad
\chi''_{\rm
e}=\frac{N|\wp_{ab}|^{2}\Delta^{2}\gamma_{1}}{\epsilon_{0}\hbar\left[\left(\Delta^{2}-\frac{\Omega_{\rm
c}^{\ast}\Omega_{\rm
c}}{4}\right)^{2}+\Delta^{2}\gamma_{1}^{2}\right]}. \label{parts}
\end{equation}
For the present, we focus our attention only on the low-absorption
EIT medium. This requires that the imaginary part $\chi''_{\rm e}$
is negligibly small (or $\chi''_{\rm e}\ll |1+\chi'_{\rm e}|$). In
an ordinary EIT experiment, the Rabi frequency, $\Omega_{\rm c}$,
of the coupling laser often has the same order of magnitude of
$\gamma_{1}$, {\it e.g.}, $10^{7}\sim 10^{8}$
s$^{-1}$\cite{Li,Schmidt}. Thus, it follows from Eqs.
(\ref{parts}) that for a low-absorption EIT medium, the frequency
detuning of the probe field should be much less than the Rabi
frequency of the coupling laser, {\it i.e.}, $\Delta
\ll\Omega_{\rm c}$. To achieve the negative electric permittivity,
{\it i.e.}, $\chi'_{\rm e}< -1$, according to Eqs. (\ref{parts}),
the quantities $\Delta$, $\Omega_{\rm c}$, $\gamma_{1}$,
$\wp_{ab}$ and $N$ should agree with the following restriction
condition
\begin{equation}
\zeta \Delta \left(\Delta^{2}-\frac{\Omega_{\rm
c}^{\ast}\Omega_{\rm
c}}{4}\right)+\left(\Delta^{2}-\frac{\Omega_{\rm
c}^{\ast}\Omega_{\rm c}}{4}\right)^{2}+\Delta^{2}\gamma_{1}^{2}< 0
\label{inequality}
\end{equation}
with $\zeta=N|\wp_{ab}|^{2}/\left(\epsilon_{0}\hbar\right)$. Since
$\Delta \ll\Omega_{\rm c}$, one can ignore the $\Delta^{2}$ term
in inequality (\ref{inequality}), and then obtain $\Delta
>{\Omega_{\rm c}^{\ast}\Omega_{\rm c}}/\left({4\zeta}\right)$ from the inequality (\ref{inequality}).
Thus, the requirement for the realization of negative permittivity
($\epsilon_{\rm r}\simeq 1+\chi'_{\rm e}<0$) near the probe
frequency in a three-level atomic system is $ \Omega_{\rm c}\gg
\Delta
>{\Omega_{\rm c}^{\ast}\Omega_{\rm c}}/{4\zeta}$.
It should be noted that the condition $\Omega_{\rm c}\gg
{\Omega_{\rm c}^{\ast}\Omega_{\rm c}}/\left({4\zeta}\right)$
imposes a restriction on the atomic number density $N$ of the EIT
medium, {\it i.e.}, $ N\gg {\epsilon_{0}\hbar |\Omega_{\rm
c}^{\ast}|}/{4|\wp_{ab}|^{2}}$. This, therefore, means that in
order to realize the low-absorption negative permittivity, one
should choose the dense-gas EIT media. For a rough discussion, we
choose the values for the dipole matrix elements and decay rates,
which are typical for transitions in hyperfine-split Na D
lines\cite{Cowan}: $\wp_{ab}=1.2\times 10^{-31}$ D,
$\gamma_{1}=1.2\times 10^{8}$ s$^{-1}$. In this case, $\zeta\simeq
10^{-17}N$ m$^{3}$s$^{-1}$. If the Rabi frequency of coupling
laser takes the value of $10^{8}$ s$^{-1}$\cite{Li,Schmidt}, the
detuning $\Delta$ of the probe laser should be larger than
$10^{33}/N$ m$^{-3}$s$^{-1}$ in accordance with the condition
$\Delta
>{\Omega_{\rm c}^{\ast}\Omega_{\rm c}}/({4\zeta})$, and then the EIT medium will exhibit a
negative permittivity. As far as the low absorption is concerned,
the atomic number density $N$ should be much larger than $10^{24}$
m$^{-3}$, according to the requirement $ N\gg {\epsilon_{0}\hbar
|\Omega_{\rm c}^{\ast}|}/{4|\wp_{ab}|^{2}}$. It follows that the
larger is the atomic density, the more negative is the
permittivity obtained at nearly resonant frequency with low
absorption.

As for the treatment for the negative permeability in the EIT
medium, the only task left to us is to discuss the parameter
$\varsigma^{2}$. As an illustrative example, here we are concerned
only with the case of the large negative permittivity ({\it i.e.},
$\chi'_{\rm e}\ll -1$). In this case, $\varsigma\rightarrow
i\left(m_{cb}^{\ast}/\wp_{ab}^{\ast}c\right)\sqrt{\Omega_{\rm
c}^{\ast}/\Omega_{\rm c}}\sqrt{\zeta/\Delta}$. This, therefore,
means that the restriction for realizing the negative permeability
can be rewritten as $
\left(\frac{m_{cb}^{\ast}}{\wp_{ab}^{\ast}c}\right)^{2}\left(\frac{\Omega_{\rm
c}^{\ast}}{\Omega_{\rm c}}\right)\frac{\zeta}{\Delta}\gg 4$. Note
that in an atomic system,
${m_{cb}^{\ast}}/\left({\wp_{ab}^{\ast}c}\right)\sim 10^{-2}$. If
we choose $\Omega_{\rm c}^{\ast}=\Omega_{\rm c}$, for a tentative
consideration, the above condition can be rewritten as
$\Delta\ll\zeta/(4\times 10^{4})$, {\it i.e.}, $\Delta\ll
10^{-21}N$ m$^{3}$s$^{-1}$.

In conclusion, for a typical EIT experiment in which
$\gamma_{1}\sim\Omega_{\rm c}\simeq 10^{8}$ s$^{-1}$, in order to
obtain the negative permittivity, the frequency detuning $\Delta$
of the probe laser should satisfy $\Delta>10^{33}/N$
m$^{-3}$s$^{-1}$; in order to obtain the negative permeability,
$\Delta$ should satisfy $\Delta\ll 10^{-21}N$ m$^{3}$s$^{-1}$; in
order to obtain the low absorption for the probe light, the two
requirements $\Delta\ll \Omega_{\rm c}$ and $N\gg 10^{24}$
m$^{-3}$ should be satisfied. Thus, it is shown that only when the
three-level atomic system is dense gas, the atomic density $N$ of
which is larger than $10^{27}$ m$^{-3}$ and the frequency detuning
$\Delta$ is approximately or less than $10^{6}$ s$^{-1}$ will such
a three-level EIT medium exhibit a low-absorption negative optical
refractive index. The behaviors of permittivity, permeability and
absorption for the probe laser are shown in Figs. 2 and 3.

\section{PHYSICAL SIGNIFICANCE OF the three-level EIT-BASED REALIZATION OF NEGATIVE REFRACTIVE INDEX}
We proposed a novel scheme of EIT-based realization of left-handed
media. Such a scenario of achieving the negative refractive index
with EIT atomic gas may have three main advantages: \\ (i) can
realize the {\it isotropic} left-handed media; \\ (ii) can lead to
a controllable manipulation of negative refractive index by an
external field (the coupling laser); \\ (iii) can obtain the
visible and infrared frequency band where the permittivity and the
permeability are simultaneously negative. For these reasons, we
think that the present scheme for realizing the EIT left-handed
media deserves further investigation both theoretically and
experimentally.
\\ \\

\textbf{Acknowledgements}   This project is supported by the
National Natural Science Foundation (NSF) of China under Project
No. $90101024$ and $60378037$. The subject of this paper is one of
the original ideas in our national 973 programme application [the
National Basic Research Program (973) of China (Project No.
2004CB719805)] submitted in April, 2004, in which Prof. Jian-Ying
Zhou of Zhongshan University was also involved.

\newpage

\section*{Figure captions}

Fig. 1. The schematic diagram of the $\Lambda$-type three-level
EIT atomic system. The electric-dipole transition
($|a\rangle$-$|b\rangle$) and the magnetic-dipole transition
($|c\rangle$-$|b\rangle$) will possibly yield the negative
electric permittivity and the magnetic permeability at probe
frequency, respectively.
\\ \\

Fig. 2. The real and imaginary parts of the permittivity and
permeability. It is shown that in the case of small frequency
detuning, the permittivity has a negative real part and a very
small imaginary part if the frequency detuning $\Delta$ is greater
than $2.3\times 10^{-3}\gamma_{1}$. When the frequency detuning
ranges from $2.3\times 10^{-3}\gamma_{1}$ to $7.7\times
10^{-3}\gamma_{1}$, the dense EIT atomic gas will exhibit a
negative permeability. In the band of frequency detuning
$[2.3\times 10^{-3}\gamma_{1}, 4.6\times 10^{-3}\gamma_{1}]$, the
imaginary part of permeability is negligibly small. In the
detuning range $[4.6\times 10^{-3}\gamma_{1}, 7.7\times
10^{-3}\gamma_{1}]$, however, the imaginary part of permeability
increases significantly. Here, $N=10^{27}$ m$^{-3}$,
$\gamma_{1}=\Omega_{\rm c}\simeq 10^{8}$ s$^{-1}$,
$\wp_{ab}=1.2\times 10^{-31}$ D and
${m_{cb}^{\ast}}/\left({\wp_{ab}^{\ast}c}\right)=10^{-2}$.
\\ \\

Fig. 3. The absorption coefficient ($\alpha=-2\pi {\rm
Im}\{\sqrt{\epsilon_{\rm r}\mu_{\rm r}}\}$). It is shown that the
probe laser will experience a small amplification in the detuning
range $[0, 2.3\times 10^{-3}\gamma_{1}]$, where both the
permittivity and permeability are positive numbers. However, in
the detuning range $[2.3\times 10^{-3}\gamma_{1}, 7.7\times
10^{-3}\gamma_{1}]$ for the EIT atomic gas to be a left-handed
medium, such an EIT medium is an absorptive material. For the case
of $\Delta<4.6\times 10^{-3}\gamma_{1}$, it is a low-absorptive
medium, while for the case of $\Delta>4.6\times
10^{-3}\gamma_{1}$, it is a high-absorptive one, for the
permeability has a large imaginary part.

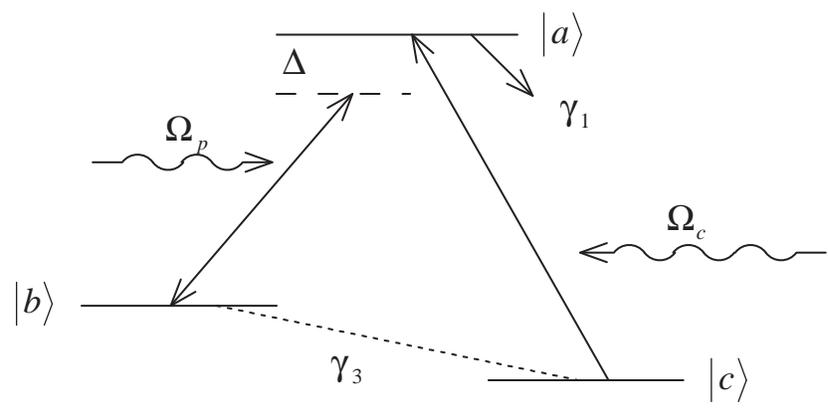
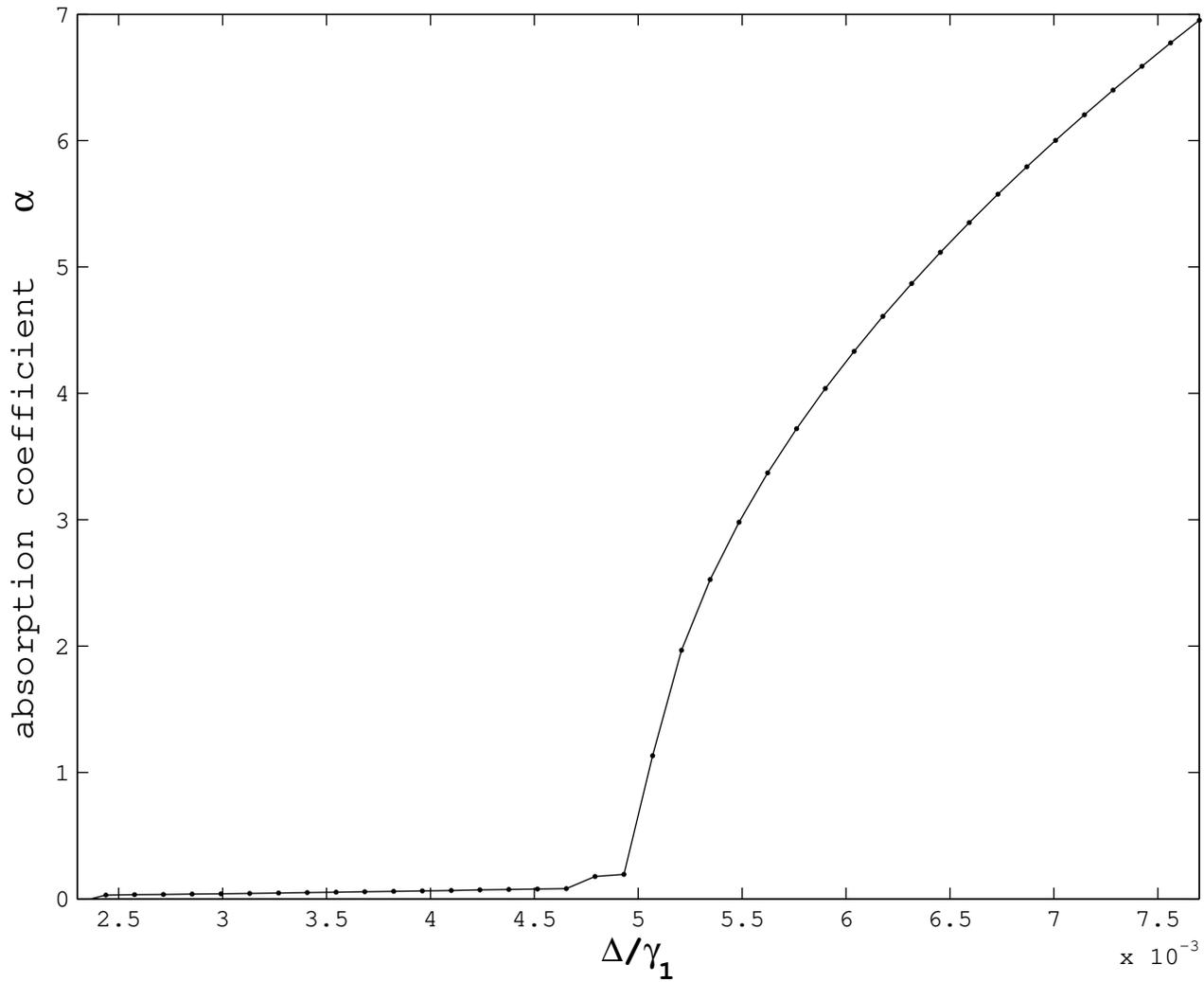
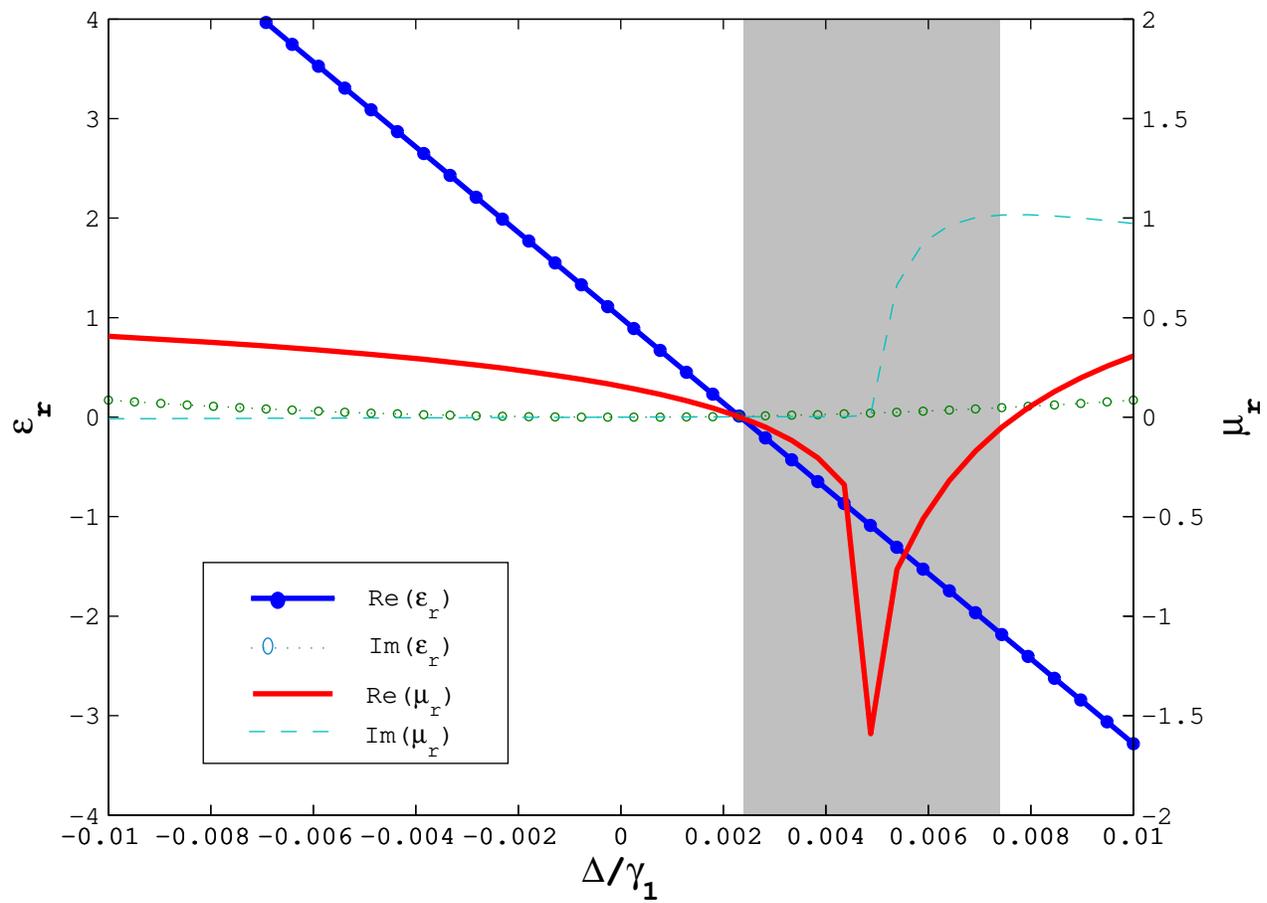
\end{document}